%% file: P040055.tex
\documentclass[twocolumn,superscriptaddress,showpacs,preprintnumbers,amsmath,amssymb,aas_macros]{revtex4}

\usepackage{graphicx}
\usepackage{amssymb}
\usepackage{natbib}
\newcommand\ligodoc{P040055-01-Z}

\begin{document}
\preprint{LIGO-\ligodoc}

\title{Search for gravitational wave radiation associated with the pulsating tail of the SGR $\bf 1806-20$ hyperflare of 27 December 2004 using LIGO}

\begin{abstract}
We have searched for Gravitational Waves (GWs) associated with the SGR $1806-20$ hyperflare of 27 December 2004. This event, originating from a Galactic neutron star, displayed exceptional energetics. Recent investigations of the X-ray light curve's pulsating tail revealed the presence of Quasi-Periodic Oscillations (QPOs) in the $30 - 2000 ~ \mathrm{Hz}$ frequency range, most of which coincides with the bandwidth of the LIGO detectors. These QPOs, with well-characterized frequencies, can plausibly be attributed to seismic modes of the neutron star which could emit GWs. Our search targeted potential quasi-monochromatic GWs lasting for tens of seconds and emitted at the QPO frequencies. We have observed no candidate signals above a pre-determined threshold and our lowest upper limit was set by the $92.5 ~ \mathrm{Hz}$ QPO observed in the interval from $150 ~ \mathrm{s}$ to $260 ~ \mathrm{s}$ after the start of the flare. This bound corresponds to a ($90\%$ confidence) root-sum-squared amplitude $\mathrm{h}_\mathrm{rss-det}^{90\%} = 4.5 \times 10^{-22} ~ \mathrm{strain ~ Hz}^{-1/2}$ on the GW waveform strength in the detectable polarization state reaching our Hanford (WA) $4 ~ \mathrm{km}$ detector. We illustrate the astrophysical significance of the result via an estimated characteristic energy in GW emission that we would expect to be able to detect. The above result corresponds to $7.7 \times 10^{46} ~ \mathrm{erg} ~ ( = 4.3 \times 10^{-8} \mathrm{M}_{\odot} \mathrm{c}^2)$, which is of the same order as the total (isotropic) energy emitted in the electromagnetic spectrum. This result provides a means to probe the energy reservoir of the source with the best upper limit on the GW waveform strength published and represents the first broadband asteroseismology measurement using a GW detector.
\end{abstract}

\include{LSCauthors}



\maketitle

\section{Introduction}

Soft Gamma-ray Repeaters (SGRs) are objects that emit short-duration X and gamma-ray bursts at irregular intervals (see \cite{woods04} for a review). These recurrent bursts generally have durations of the order of $\sim 100 ~ \mathrm{ms}$ and luminosities in the $10^{39} - 10^{42} ~ \mathrm{erg/s}$ range. At times, though rarely, these sources emit giant flares lasting hundreds of seconds (see for example \cite{mazets79,hurley99,hurley05}) with peak electromagnetic luminosities reaching $10^{47} ~ \mathrm{erg/s}$ \cite{woods06}. Pulsations in the light curve tail reveal the neutron star spin period.

Quasi-Periodic Oscillations (QPOs) \cite{israel05, watts06, strohmayer06, levin06a, levin06b} in the pulsating tail of giant flares were first observed for the 27 December 2004 event of SGR $1806-20$ by the {\it Rossi X-Ray Timing Explorer (RXTE)} and {\it Ramaty High Energy Solar Spectroscopic Imager (RHESSI)} satellites \cite{israel05,watts06, strohmayer06}. Prompted by these observations, the RXTE data from the SGR $1900+14$ giant flare of 27 August 1998 was revisited \cite{strohmayer05}. Transient QPOs were found in the light curve pulsating tail at similar frequencies to the SGR $1806-20$ event, suggesting that the same fundamental physical process is likely taking place.

\begin{table*}
\begin{center}
\begin{tabular}{cccccc}

{\bf Observation} & {\bf Frequency} & {\bf FWHM} & {\bf Period} & {\bf Satellite} & {\bf References}\\

 & {\bf [Hz]}  & {\bf [Hz]} & {\bf [s]} &  & \\

\hline \hline \\

a &  17.9 $\pm$ 0.1 & 1.9 $\pm$ 0.2 & 60-230 & RHESSI & \cite{watts06}\\ \\

b &  25.7 $\pm$ 0.1 & 3.0 $\pm$ 0.2 & 60-230 & RHESSI & \cite{watts06}\\ \\

c &  29.0 $\pm$ 0.4 & 4.1 $\pm$ 0.5 & 190-260 & RXTE  & \cite{strohmayer06}\\ \\

d &  92.5 $\pm$ 0.2 & $1.7^{+ 0.7}_{-0.4}$ & 170-220 & RXTE & \cite{israel05}\\

e & "               & "                    & 150-260 & "    & \cite{strohmayer06}\footnote{Ref. \cite{strohmayer06} makes an adjustment to the observation period of Ref. \cite{israel05}} \\

f & 92.7 $\pm$ 0.1 & 2.3 $\pm$ 0.2 & 150-260 & RHESSI & \cite{watts06}\\

g &  92.9 $\pm$ 0.2 & 2.4 $\pm$ 0.3 & 190-260 & RXTE & \cite{strohmayer06}\\ \\

h &  150.3 $\pm$ 1.6 & 17 $\pm$ 5 & 10-350 & RXTE & \cite{strohmayer06}\\ \\

i &  626.46 $\pm$ 0.02 & 0.8 $\pm$ 0.1 & 50-200 & RHESSI & \cite{watts06} \\

l &  625.5 $\pm$ 0.2 & 1.8 $\pm$ 0.4 & 190-260 & RXTE & \cite{strohmayer06}\\ \\

m & 1837 $\pm$ 0.8 & 4.7 $\pm$ 1.2 & 230-245 & RXTE & \cite{strohmayer06}\\ \\

\hline \hline

\end{tabular}
\end{center}
\caption{Summary of the most significant QPOs observed in the pulsating tail of SGR $1806-20$ during the 27 December 2004 hyperflare (from Ref. \cite{strohmayer06}). The period of observation for the QPO transient is measured with respect to the flare peak, the frequencies are given from the Lorenzian fits of the data and the width corresponds to the Full-Width-at-Half-Maximum (FWHM) of the given QPO band.}
\label{tab:tableofqpos}
\end{table*}

Several characteristics of SGRs can be explained by the {\it magnetar} model \cite{duncan92}, in which the object is a neutron star with a high magnetic field ($B \sim 10^{15} ~ \mathrm{G}$). In this model the giant flares are generated by the catastrophic rearrangement of the neutron star's crust and magnetic field, a {\it starquake} \cite{schwartz05,palmer05}.

It has been suggested that the star's seismic modes, excited by this catastrophic event, might drive the observed QPOs \cite{duncan98, israel05, watts06, strohmayer06}, which leads us to investigate a possible emission of Gravitational Waves (GWs) associated with them. There are several classes of non-radial neutron star seismic modes with characteristic frequencies in the $\sim 10 - 2000 ~ \mathrm{Hz}$ range \cite{mcdermott88}. Toroidal modes of the neutron star crust are expected to be excited by large crustal fracturing (see \cite{israel05, watts06, strohmayer06, piro05}), though these modes may be poor GW emitters. However, crust modes could magnetically couple to the core's modes, possibly generating a GW signal accessible with today's technology (see \cite{thorne83,glampedakis06a,glampedakis06b}). Other modes with expected frequencies in the observed range are crustal interface modes, crustal spheroidal modes, crust/core interface modes or perhaps p-modes, g-modes or f-modes. The latter should, in theory, be stronger GW emitters (see for example \cite{andersson97, andersson02}).

In addition, it has been noted \cite{owen05} that a normal neutron star can only store a crustal elastic energy of up to $\sim 10^{44} ~ \mathrm{erg}$ before breaking. An alternative to the conventional neutron star model, that of a solid quark star, has also been proposed in several versions \cite{xu03, owen05, xu06, mannarelli07}. In this case an energy of $\sim 10^{46} ~ \mathrm{erg}$ (as observed for this flare) is feasible, and thus the mechanical energy in the GW-emitting crust oscillations could be comparable to the energy released electromagnetically. This was also noted by Horvath \cite{horvath05}, who in addition estimated that LIGO might be able to detect a GW burst of comparable energy to the electromagnetic energy (this was before the QPOs were discovered.)

The exceptional energetics of the SGR $1806-20$ hyperflare \cite{hurley05, palmer05}, the close proximity of the source \cite{hurley05, cameron05, corbel04, mcclure05} and the availability of precisely measured QPO frequencies and bandwidths \cite{israel05, watts06, strohmayer06} made SGR $1806-20$ attractive for study as a possible GW emitter.

In this paper we make use of the LIGO Hanford (WA) $4 ~ \mathrm{km}$ detector (H1), the only LIGO detector collecting low noise data at the time of the flare, to search for or to place an upper bound on the GW emission associated with the observed QPO phenomena of SGR $1806-20$. At the time of the event the GEO600 detector was also collecting data. However, due to its significantly lower sensitivity at the frequencies of interest, it was not used in this analysis.

As will be shown, the $92.5 ~ \mathrm{Hz}$ QPO upper bounds can be cast into a characteristic GW energy release in the $\sim 8 \times 10^{46} -~ 3 \times 10^{47} ~ \mathrm{erg} ~ ( \sim 4 \times 10^{-8} -~ 2 \times 10^{-7} ~ \mathrm{M}_{\odot} \mathrm{c}^2)$ range. This energy approaches the total energy emitted in the electromagnetic spectrum and offers the opportunity to explore the energy reservoir of the source. In the event of a similar Galactic hyperflare coinciding with LIGO's fifth science run (S5), the energy sensitivity involved at $\sim 100 ~ \mathrm{Hz}$ would probe the $\sim 2 \times 10^{45} ~ \mathrm{erg} ~ ( \sim 10^{-9} ~ \mathrm{M}_{\odot} \mathrm{c}^2)$ regime.

\section{Satellite observations}

SGR $1806-20$ is a Galactic X-ray star thought to be at a distance in the $6$ to $15 ~ \mathrm{kpc}$ range \cite{hurley05,cameron05,corbel04,mcclure05}. The total (isotropic) electromagnetic flare energy for the 27 December 2004 record flare was measured to be $\sim 10^{46} ~ \mathrm{ergs}$ \cite{hurley05,palmer05} assuming a distance of $10 ~ \mathrm{kpc}$.

QPOs in the pulsating tail of the SGR $1806-20$ hyperflare were first observed by Israel {\it et al.}\cite{israel05} using RXTE, and revealed oscillations centered at $\sim 18$, $\sim 30$ and $\sim 92.5 ~ \mathrm{Hz}$. Using RHESSI, Watts and Strohmayer \cite{watts06} confirmed the QPO observations of Israel {\it et al.} revealing additional frequencies at $\sim 26 ~ \mathrm{Hz}$ and $\sim 626.5 ~ \mathrm{Hz}$ associated with a different rotational phase. Closer inspection of the RXTE data by Strohmayer and Watts \cite{strohmayer06} revealed a richer presence of QPOs, identifying significant components at $\sim 150$ and $\sim 1840 ~ \mathrm{Hz}$ as well. Table \ref{tab:tableofqpos} is taken from Ref. \cite{strohmayer06} and summarizes the properties of the most significant QPOs detected in the X-ray light curve tail of the SGR $1806-20$ giant flare.

\section{The LIGO detectors}

The Laser Interferometer Gravitational Wave Observatory (LIGO) \footnote{http://www.ligo.caltech.edu} consists of three detectors, two located at Hanford, WA (referred to as H1 and H2) and a third located in Livingston, LA (referred to as L1). Each of the detectors consists of a long-baseline interferometer in a Michelson configuration with Fabry-Perot arms (see Ref. \cite{ligobservatory} for details). The passage of a GW induces a differential arm length change $\Delta L$ which is converted to a photocurrent by a photosensitive element monitoring the interference pattern of the detector. This electrical signal is then amplified, filtered and digitized at a rate of $16384 ~ \mathrm{Hz}$ to produce a time series which we refer to as the GW channel.

To calibrate the GW channel in physical units, the interferometer response function is frequently measured by generating known differential arm length changes. The uninterrupted monitoring of the response function is ensured with the addition of continuous sinusoidal excitations referred to as {\it calibration lines}.

The interferometer sensitivity  to $\Delta L$ enables to measure a strain $h$ defined as
\begin{equation}
\label{eq:strain}
h = \frac{\Delta L}{L}
\end{equation}
where $L$ denote the mean of the two arm lengths. The target frequency range of interest is the audio band with frequencies in the $50 ~ \mathrm{Hz}$ to $7 ~ \mathrm{kHz}$ range.

LIGO has dedicated science runs when good and reliable coincidence data is available, alternating with periods of commissioning to improve the sensitivity of the instrument. In order to cover times when an astrophysically notable event might occur, such as the 27 December 2004 event of this analysis, data from times when commissioning activities do not disable the machine is archived by a program referred to as {\it Astrowatch} \cite{astrowatch}. Due to the nature of the time period, the detector's configuration was continuously evolving and was not as well characterized as the dedicated science runs. On the other hand, there was a deliberate attempt to place the interferometers in a high-sensitivity configuration compatible with the commissioning modifications of the epoch.

At the time of this event two of the LIGO detectors were undergoing commissioning in preparation for the fourth science run (S4). Only data from H1 is available for the analysis of this event.

\begin{figure*}
  \includegraphics[width=5in]{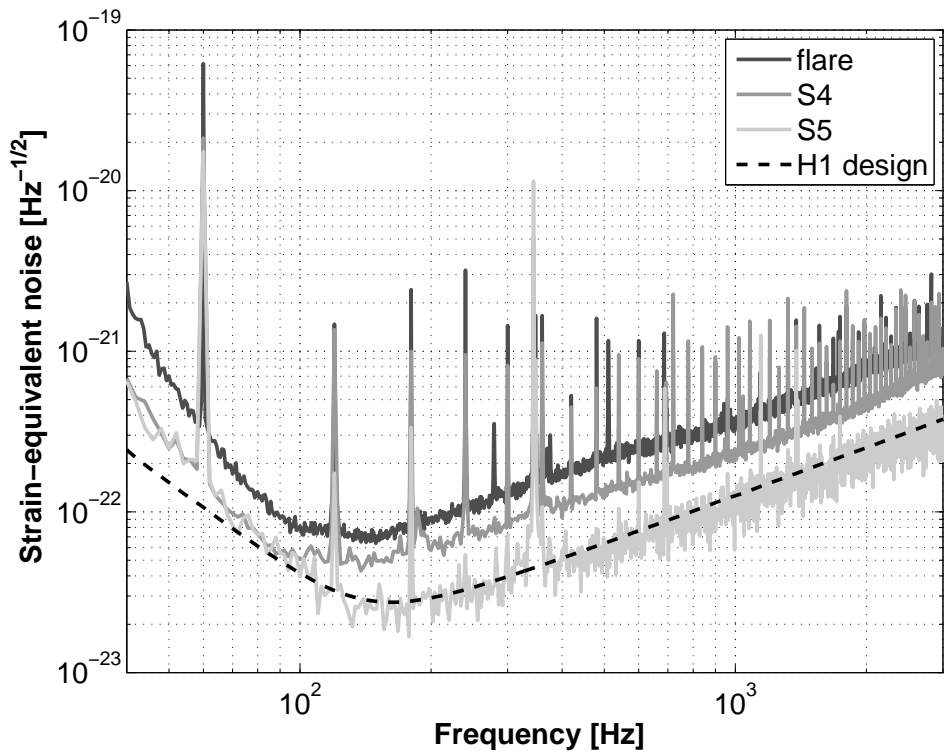}\\
  \caption{The strain-equivalent sensitivity of the H1 detector at the time of the hyperflare, the fourth and fifth science runs (S4, S5), and its design sensitivity.}
  \label{fig:fig_sensitivity}
\end{figure*}

Figure \ref{fig:fig_sensitivity} plots the best strain-equivalent noise spectra of H1 during the S4 and S5 data-taking periods (light gray curves). The average noise spectra at the time of the flare is shown by the dark gray curve and the dashed line describes the design sensitivity.

\section{Data Analysis}

This analysis relies on an {\it excess power} search \footnote{The source code used for this analysis can be found at the LSC Data Analysis Software web site http://www.lsc-group.phys.uwm.edu/daswg/projects/matapps.html with path searches/burst/QPOcode (release-1$\_$0) under the matapps tree.}, variants of which are described in Ref. \cite{flanagan98a, flanagan98b, anderson01}. In this analysis we compare time-frequency slices at the time of the observations with neighboring ones. The algorithm used analyzes a single data stream at multiple frequency bands and can easily be expanded to handle coincident data streams from multiple detectors. The trigger provided for the analysis corresponds to the flare's X-ray peak as provided by the GRB Coordinate Network (GCN) reports 2920 \footnote{http://gcn.gsfc.nasa.gov/gcn3/2920.gcn3} and 2936 \footnote{http://gcn.gsfc.nasa.gov/gcn3/2936.gcn3} at time corresponding to 21:30:26.65 UTC of 2004-12-27.

In the absence of reliable theoretical models of GW emission from magnetars, we keep the GW search as broad and sensitive as possible. The search follows the QPO signatures observed in the electromagnetic spectrum both in frequency and time interval. In particular, we measure the power (in terms of detector strain) for the intervals at the observed QPO frequencies (as shown in Tab. \ref{tab:tableofqpos}) for a given bandwidth (typically $10 ~ \mathrm{Hz}$) and we compare it to the power measured in adjacent frequency bands not related to the QPO. The excess power is then calculated for each time-frequency volume of interest.

Although QPOs are not observed in X-rays until some time after the flare, the magnetar model suggests that the seismic modes would be excited at the time of the flare itself. For this reason, we also search for GW emission associated with the proposed seismic modes from the received trigger time of the event. In addition, we chose to examine arbitrary selected frequency bands, referred to as control bands, whose center frequency is set to twice the QPO frequency and processed identically to the QPO bands. This allowed us to cover a wider range of the detector's sensitivity while allowing the reader the flexibility to estimate the sensitivity to low significance QPOs not addressed here (see Ref. \cite{strohmayer06}) as well as future observations/exotic models of GW emission yet to come.

Another aspect of the satellite observations is the quasi-periodic nature of the emitted electromagnetic waveform with a possible slow drift in frequency. Since there is no knowledge of the GW waveforms that would be associated with this type of event, we tune our search algorithm to be most sensitive to long quasi-periodic waveforms with fairly narrow bandwidths while short bursts are strongly discriminated against. The waveform set used in testing the sensitivity of the algorithm by adding simulated data in the analysis software is chosen in line with this argument.

\subsection{Pipeline}
\label{sec:pipeline}

\begin{figure}
  \includegraphics[width=2in]{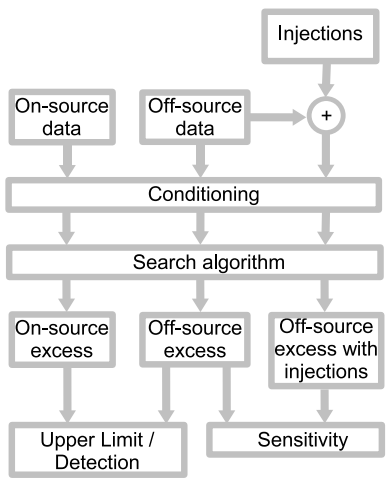}\\
  \caption{A block diagram of the analysis sketching the signal flow.}\label{fig:pipeline}
\end{figure}

A block diagram of the analysis pipeline is shown in Fig. \ref{fig:pipeline} where the Gamma-ray bursts Coordinates Network (GCN) reports provide the trigger for the analysis. The {\it on} and {\it off}-source data regions are then selected where the former corresponds to the QPO observation periods, as shown in Tab. \ref{tab:tableofqpos}. The {\it off}-source data region begins at the end of the six minute long QPO tail (set to $400 ~ \mathrm{s}$ after the flare peak) lasting to ten minutes prior to the end of the stable H1 lock stretch for a total of $\sim 2 ~ \mathrm{h}$ of data.

The on-source region consists of a single segment. This segment either starts at the moment of the flare ($t_\mathrm{start} = t_0$) or at the beginning of the QPO observation ($t_\mathrm{start} = t_{\mathrm{qpo}}$) and lasts until the end of the observation ($t_\mathrm{end}$). The off-source region consists of numerous non-overlapping segments, each of duration $\Delta t = t_\mathrm{end} - t_\mathrm{start}$.

To provide an estimate of the search sensitivity, an arbitrary simulated gravitational waveform can be added (or {\it injected}) to each off-source data segment. All of the segments ({\it on}- or {\it off}-source) are processed identically. In the procedure described by the conditioning block, the data is band-pass filtered to select the three frequency bands of interest: the QPO band as shown in Tab. \ref{tab:tableofqpos} and the two adjacent frequency bands. Using the interferometer response function at the time of the event, the data is calibrated into units of strain and a data-quality procedure, as described below, is applied to the data set.

After the conditioning procedure is complete, the data stream is pushed through the search algorithm, which computes the power in each segment for the three frequency bands of interest and then the excess power in the segment. Finally, on- and off-source excesses are compared and in the case of no significant on-source signals, the standard Feldman-Cousins \cite{feldman98} statistical approach is used to place an upper limit based on the loudest signal.

The data processing can be validated against analytical expectations by replacing the off-source region with simulated data.

\subsection{Data conditioning}

The conditioning procedure consists of zero-phase filtering of the data with three different band-pass Butterworth filters. The first band-pass filters the data around the QPO frequency of interest with a predefined bandwidth. This bandwidth depends on the observed QPO width (see Tab. \ref{tab:tableofqpos}) and on the fact that the QPOs have been observed to evolve in frequency. For the QPOs addressed here, the bandwidth is set to $10 ~ \mathrm{Hz}$ (well above the measured FWHM shown in Tab. \ref{tab:tableofqpos} with the exception of the $150.3 ~ \mathrm{Hz}$ oscillation where the bandwidth was set to the measured FWHM, $17 ~ \mathrm{Hz}$.

The bandwidth for the control bands is also set to $10 ~ \mathrm{Hz}$ which is still above twice the measured FWHM. An exception to this is the $150.3 ~ \mathrm{Hz}$ second harmonic which is within one Hz away from the fifth harmonic of the $60 ~ \mathrm{Hz}$ power line. The bandwidth in this case is set to twice the measured FWHM ($2 * 17 ~ \mathrm{Hz} = 34 ~ \mathrm{Hz}$) but a $4 ~ \mathrm{Hz}$ wide notch at $300 ~ \mathrm{Hz}$ is included to suppress the significant sensitivity degradation provided by the line. For this reason, the effective bandwidth is $30 ~ \mathrm{Hz}$.

The data is also filtered to select the two adjacent frequency bands with identical bandwidths of the chosen QPO band. Using the adjacent frequency bands allows us to discriminate against common non-stationary broad-band noise, thereby increasing the search sensitivity, as will be described in Sec. \ref{sec:searchalgorithm}.

A gap between frequency bands was introduced for some of the QPO frequencies in order to minimize the power contribution of known instrumental lines. Furthermore, $60 ~ \mathrm{Hz}$ harmonics which landed in the bands of interest were strongly suppressed using narrow notch filters.

The three data streams are calibrated in units of strain using a transfer function which describes the interferometer response to a differential arm length change.

The conditioning procedure ends with the identification of periods of significant sensitivity degradation. These periods are selected by monitoring the power in each of the three frequency bands in data segment durations, or {\it tiles}, $125 ~ \mathrm{ms}$ and $1 ~ \mathrm{s}$ long. If the power is above a set threshold in any of the three bands, the tile in question identifies a period of noise increase. This abrupt power change in a second-long time frame (or less) does not correspond to a GW candidate lasting tens to hundreds of seconds long. For this reason, the full data set contained in the identified tile is disregarded and short-duration GW bursts, not among the targeted signals, would be excluded by this analysis.

To set a particular threshold we first determined the variance of the resulting power distribution which was calculated by removing outliers iteratively. As will be described in Sec. \ref{sec:sensitivity}, we used $2 \sigma$, $3 \sigma$ and $4 \sigma$ cuts and we injected different waveform families to optimize the search sensitivity.

\subsection{The search algorithm}
\label{sec:searchalgorithm}

The algorithm at the root of the search consists of taking the difference in power between a band centered at a frequency $f_\mathrm{qpo}$ and the average of the two frequency bands adjacent to the QPO frequency band, also of bandwidth $\Delta f$, typically centered at $f_{\pm} = f_\mathrm{qpo} \pm \Delta f$.

After band-pass filtering, we are left with three channels for each QPO: $c_\mathrm{qpo}(t)$, $c_+(t)$, and $c_-(t)$. The power for the QPO interval is for each of these channels:
\begin{equation}
{\cal P}_{\mathrm{qpo},\pm} = \int_{t_\mathrm{start}}^{t_\mathrm{end}} ( c_{\mathrm{qpo},\pm} )^2 dt
\end{equation}
where tiles that were vetoed are excluded from the integral. The excess power is then defined as
\begin{equation}
\label{eq:excesspower}
\Delta {\cal P} = {\cal P}_\mathrm{qpo} - {\cal P}_{\mathrm{avg}}
\end{equation}
where ${\cal P}_{\mathrm{avg}} = ({\cal P}_{\mathrm{+}} + {\cal P}_{\mathrm{-}} ) / 2$ is the average of the adjacent bands. We refer to the resulting set of $\Delta {\cal P}$ calculated over the off-source region as the {\it background} while the on-source region provides a single excess power measurement of duration $\Delta t$ for the period from $t_\mathrm{start}$ to $t_\mathrm{end}$.

\section{Sensitivity of the search}
\label{sec:sensitivity}

In order to estimate the sensitivity of the search, different sets of more or less astrophysically-motivated waveforms, or in some cases completely {\it ad-hoc} waveforms, are injected in the off-source region and the resulting excess power is computed.

The strength of the injected strain (at the detector) $h_\mathrm{det}(t)$ is defined by its {\it root-sum-square} (rss) amplitude, or
\begin{equation}
h_\mathrm{rss-det} = \sqrt{ \int_{t_1}^{t_1 + \Delta t} | h_\mathrm{det}(t) |^2 \mathrm{dt}}
\end{equation}
integrated over the interval $\Delta t$, as described in Sec. \ref{sec:pipeline}, where $t_1$ indicates the start of a segment in the background region. The search sensitivity to a particular waveform, $h_\mathrm{rss-det}^\mathrm{sens}$, is defined as the injected amplitude $h_\mathrm{rss-det}$ such that $90\%$ of the resulting $\Delta {\cal P}$ is above the off-source median. This choice of definition provides a {\it characteristic} waveform strength which, on average, should not be far from a $90\%$ upper bound.

\begin{figure*}
  \includegraphics[width=5.in]{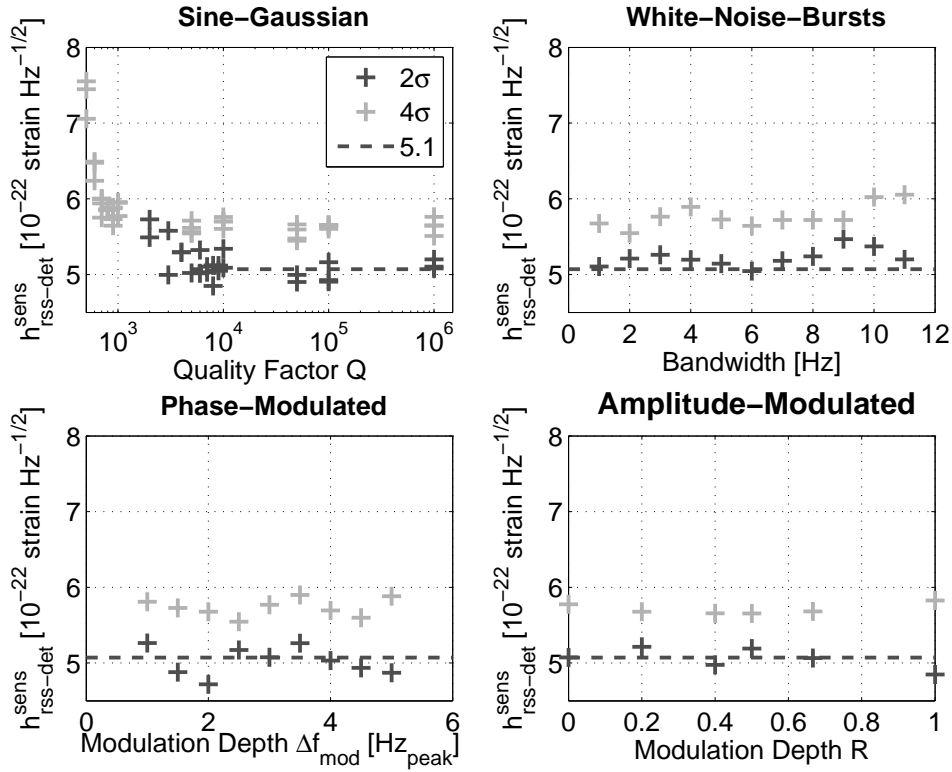}\\
  \caption{Search sensitivity to different waveform families and for different data quality cuts. The cuts are relative to the off-source RMS distribution calculated in segments $125 ~ \mathrm{ms}$ long and for $2 \sigma$ cuts (dark gray crosses) and $4 \sigma$ cuts (light gray crosses). Top left: SG waveforms injections as a function of quality factor Q varied from $Q = 600$ to $Q = 10^6$. Dashed line represents the average sensitivity ($5.1 \times 10^{-22} ~ \mathrm{strain ~ Hz}^{-1/2}$) for injections with $Q > 5 \times 10^3$ (where the sensitivity is essentially flat) and a $2 \sigma$ cut. Top right: $40 ~ \mathrm{s}$ long WNBs waveform injections as a function of burst bandwidth ranging from $1 ~ \mathrm{Hz}$ to $11 ~ \mathrm{Hz}$. Within the parameter space explored the sensitivity is essentially constant. Bottom left and right: PM and AM waveform injections as a function of modulation depth for a modulation frequency of $100 ~ \mathrm{mHz}$.}
\label{fig:fig_sens_all}
\end{figure*}

We injected various waveform families (namely Sine-Gaussians (SG), White Noise Bursts (WNB), Amplitude (AM) and Phase Modulated (PM) waveforms) in the off-source region to quantify the sensitivity of the search to these types of waveforms. Each waveform was added directly to the raw data segments and the search sensitivity was explored as a function of the various parameters. As previously mentioned, we designed the algorithm to be sensitive to arbitrary waveforms with a preset small frequency range while discriminating against any type of short duration signals.

The result of the sensitivity study for the case of the $92.5 ~ \mathrm{Hz}$ QPO (observation {\it d} of Tab. \ref{tab:tableofqpos}) is shown in Fig. \ref{fig:fig_sens_all} where the band center frequencies, bandwidths and signal durations were set to $f_\mathrm{qpo} = 92.5 ~ \mathrm{Hz}$, $f_- = 82.5 ~ \mathrm{Hz}$, $f_+ = 102.5 ~ \mathrm{Hz}$, $\Delta f = 10 ~ \mathrm{Hz}$ and $\Delta t = 50 ~ \mathrm{s}$.

SG waveforms are parameterized as follows
\begin{equation}
\label{eq:SG}
h_\mathrm{det}(t) = A ~ \sin{(2 \pi f_\mathrm{c} t + \phi)} ~ e^{-(t-t_0)^2/ \tau^2}
\end{equation}
where $A$ is the waveform peak amplitude, $f_\mathrm{c}$ is the waveform central frequency, $Q = \sqrt{2} \pi \tau f_\mathrm{c}$ is the quality factor, $\tau$ is the $1/e$ decay time, $\phi$ is an arbitrary phase and $t_0$ indicates the waveform peak time. In the case of $Q \rightarrow \infty$ the waveform approaches the form of a pure sinusoid. The top left panel of Fig. \ref{fig:fig_sens_all} plots the search sensitivity versus the quality factor $Q$ of the injected SG waveform, indicating that the analysis is most sensitive to SG waveforms with quality factors in the range $Q \in [\sim 10^3:\infty]$. The response is also shown as a function of a $2 \sigma$ and $4 \sigma$ data quality cut on the off-source RMS distribution calculated for $125 ~ \mathrm{ms}$ long tiles. The more aggressive $2 \sigma$ cut yields significantly better results and was chosen for the $92.5 ~ \mathrm{Hz}$ QPO analysis. This band in particular is significantly more problematic than the others exhibiting a high-degree of non-stationarity as well as a relatively high glitch rate.

The decline in sensitivity as the Q decreases originates from the data quality procedure. As parameter Q takes smaller values, the waveform energy begins to concentrate in shorter time scales and the conditioning procedure identifies and removes intervals of the injection which are above threshold. In the $2 \sigma$ case, the sensitivity is relatively flat for $Q > 5 \times 10^3$ and the average value is $h_\mathrm{rss-det}^\mathrm{sens} = 5.1 \times 10^{-22} \mathrm{strain ~ Hz}^{-1/2}$ also shown in the plot by the dashed line. The corresponding waveform duration $\delta t$, defined as the interval for which the waveform amplitude is above $A/e$, is $\delta t \geq \sqrt{2} ~ Q / \pi f_\mathrm{c} \simeq 24 ~ \mathrm{s}$, appropriate for the targeted search as shown in Tab. \ref{tab:tableofqpos}.

The top right panel of Fig. \ref{fig:fig_sens_all} plots the sensitivity to a large population of $40 ~ \mathrm{s}$ long WNBs injections of bandwidths ranging from $1 ~ \mathrm{Hz}$ to $11 ~ \mathrm{Hz}$. The waveform is generated by band-passing white noise through a $2^{nd}$ order Butterworth filter with bandwidth defined at the -3dB cutoff point and burst duration set by a Tukey window. As shown in the SG case, the most aggressive $2 \sigma$ cut outperforms the $4 \sigma$ and no significant departure in sensitivity is seen for bandwidths up to $10 ~ \mathrm{Hz}$. It is worth noting that WNBs would correspond to incoherent motion of the source and may not be physical. However the purpose of this study is to quantify the robustness of the search to a variety of waveforms.

The bottom two panels of Fig. \ref{fig:fig_sens_all} plot the sensitivity to PM and AM waveforms versus modulation depth, where the modulation frequency is set to $f_\mathrm{mod} = 100 ~ \mathrm{mHz}$ for both cases. These waveforms are used to investigate QPO amplitude and frequency evolutions. For the PM case, the waveform is described as
\begin{equation}
h_\mathrm{det}(t) = A ~ \cos{ \big{(} 2 \pi f_\mathrm{c} t + k_\mathrm{mod} x(t) + \phi \big{)}}
\end{equation}
where $A$ is the waveform amplitude, $f_\mathrm{c}$ is the carrier frequency, $\phi$ is an arbitrary phase, $k_\mathrm{mod}$ is a modulation depth constant and $x(t)$ is the modulation signal
\begin{equation}
x(t) = \sin{( 2 \pi f_\mathrm{mod} t )}
\end{equation}
It can be shown that the instantaneous frequency $\hat{f}$ is
\begin{equation}
\hat{f}(t) = f_\mathrm{c} + \Delta f_\mathrm{mod} \cos(2 \pi f_\mathrm{mod} t)
\end{equation}
where $\Delta f_\mathrm{mod} = k_\mathrm{mod} ~ f_\mathrm{mod}$. From Fig. \ref{fig:fig_sens_all} the PM sensitivity is essentially constant within modulation depths in the range $\Delta f_\mathrm{mod} \in [1:5] ~ \mathrm{Hz}$.

The AM injection is parameterized as
\begin{equation} \label{eq:AM}
h_\mathrm{det}(t) = A(t) \cos{( 2 \pi f_\mathrm{c} t )}
\end{equation}
where
\begin{equation}
A(t) = \mathrm{A}_0 ~ \frac{\sin(2 \pi f_\mathrm{mod} t) - k_\mathrm{mod} }{1 + k_\mathrm{mod} }
\end{equation}
with waveform constant amplitude $\mathrm{A}_0$, $k_\mathrm{mod}$ modulation constant, and $f_\mathrm{c}$ carrier frequency. The search sensitivity to this waveform family can be expressed in terms of the modulation depth $R$ defined as
\begin{equation}
R = 1 + \frac{1 - k_\mathrm{mod} }{1 + k_\mathrm{mod} } = \frac{2}{1 + k_\mathrm{mod}}
\end{equation}

The bottom right panel of Fig. \ref{fig:fig_sens_all} plots the sensitivity of this waveform as a function of $R$. As $k_\mathrm{mod} \rightarrow \infty$, the modulation depth parameter $R \rightarrow 0$, no modulation is applied and the waveform is a sinusoid of constant amplitude. As $k_\mathrm{mod} \rightarrow 1$, the modulation depth is maximal ($R = 1$) and the amplitude $A(t)$ is also sinusoidal in nature. From Fig. \ref{fig:fig_sens_all} the AM sensitivity is essentially constant within modulation depths in the range $R \in [0:1]$. The average response to SG, as shown in the top left panel of Fig. \ref{fig:fig_sens_all}, is also shown in the other three panels for comparison.

The results shown in Fig. \ref{fig:fig_sens_all} indicate that the search sensitivity is approximately the same for all the waveforms considered.

It is also possible to estimate the theoretical search sensitivity to a sinusoidal injection. Assuming white gaussian stationary noise for the detector output, we can derive (see Ref. \cite{anderson01}) the following expression for the search sensitivity
\begin{equation}
\label{eq:sensitivity}
h_{\mathrm{rss-det}}^\mathrm{theo} \simeq 1.25 ~ S_h^{1/2}(f) ~ \big{(} \Delta f ~ \Delta t \big{)}^{1/4} \end{equation}
where $S_h^{1/2}(f)$ is the strain-equivalent amplitude spectral density of the detector noise at frequency $f$, in units of $\mathrm{strain ~ Hz}^{-1/2}$, and $\Delta f$ and $\Delta t$ are the bandwidth and duration of the segment in question, in units of $\mathrm{Hz}$ and $\mathrm{s}$. The order-of-unity factor (1.25) stems from the $90\%$ sensitivity definition as previously discussed and from taking the difference in power between bands.

Referring to Fig. \ref{fig:fig_sensitivity}, the strain sensitivity at $f = 92.5 ~ \mathrm{Hz}$ is $S_h^{1/2}(f) \simeq 9 \times 10^{-23} ~ \mathrm{strain ~ Hz}^{-1/2}$. Using $\Delta f = 10 ~ \mathrm{Hz}$ and $\Delta t = 50 ~ \mathrm{s}$, the expected sensitivity is
\begin{equation}
h_{\mathrm{rss-det}}^\mathrm{theo} \simeq 5.3 \times 10^{-22} ~ \mathrm{strain ~ Hz}^{-1/2}
\end{equation}
in good agreement with the average response of $h_\mathrm{rss-det}^\mathrm{sens} = 5.1 \times 10^{-22} ~ \mathrm{strain ~ Hz}^{-1/2}$ shown in Fig. \ref{fig:fig_sens_all}.

\section{Results}

Inspection of the on-source data segments revealed no significant departure from the off-source distribution and we cast the results of this analysis in terms of upper bounds on GW signals. These limits are found to be well below the maximum allowed upper bounds in the non-detection regime, which we refer to as non-detection threshold, assuming a continuous observation of SGR $1806-20$ and requiring an accidental rate of one event in one-hundred years (see Tab. \ref{tab:upperbounds}).

We used the unified approach of Feldman-Cousins \cite{feldman98}, which provides upper confidence limits for null results, two-sided confidence intervals for non-null results and treats confidence limits with constraints on a physical region. In view of the fact that at the time of the hyperflare event only one of the three LIGO detectors was collecting data and that the full detector diagnostic capability was not fully exploited, the lower bounds on the confidence intervals was set to zero (i.e. no detection claim based purely on the statistical analysis was allowed).

Table X of Ref. \cite{feldman98} was used to place the upper limits of this search. The excess power distribution for the off-source region of each QPO transient was parameterized with a Gaussian Probability Density function (PDF), and the mean $\mu$, standard deviation $\sigma$ and their relative errors is estimated. The on-source excess power measure and the lookup table were then used to set $90\%$ confidence intervals.

\begin{table*}
\begin{center}
\begin{tabular}{cccccc|ccc}

{\bf Observation} & {\bf Frequency} & {\bf Bandwidth} & {\bf Interval} & {\bf Duration} & {\bf Threshold$_\mathrm{\bf non-det}$ } & & $\mathrm{\bf h}_\mathrm{\bf rss-det}^{\bf 90\%}$ & \\
 & {\bf [Hz]}      & {\bf [Hz]}    & {\bf [s]}      & {\bf [s]}      & [$\mathrm{\bf 10^{-22} strain ~ Hz^{-1/2}}$] & & [$\mathrm{\bf 10^{-22} strain ~ Hz^{-1/2}}$] & \\

\hline \hline
    & & & & & & & & \\
e,f & 92.5    &  10  & 150-260   &   110   & $18.0$ & $~2.75$ & $^{ +0.47 ~ +0.70 ~ +0.16 ~ +0.77}$ &  $=4.53$ \\

g   &         &      & 190-260   &    70   & $15.7$ & $~2.90$ & $^{ +0.43 ~ +0.74 ~ +0.17 ~ +0.75}$ & $=4.67$ \\

d   &         &      & 170-220   &    50   & $14.4$ & $~5.15$ & $^{ +0.35 ~ +1.32 ~ +0.31 ~ +0.37}$ & $=7.19$ \\

    &         &      & 0-260     &   260   & $22.5$ & $~5.06$ & $^{ +1.42 ~ +1.30 ~ +0.30 ~ +2.21}$ & $=9.50$ \\
    & & & & & & & & \\

control freq. & 185.0   &   8  & 150-260   & 110     & $19.0$ & $~9.48$ & $^{ +0.51 ~ +2.43 ~ +0.57 ~ +0.27}$ & $=12.8$ \\

    &         &      & 190-260   & 70      & $17.6$ & $~8.17$ & $^{ +0.40 ~ +2.09 ~ +0.49 ~ +0.17}$ & $=11.0$ \\

    &         &      & 170-220   &    50   & $16.5$ & $~8.03$ & $^{ +0.30 ~ +2.06 ~ +0.48 ~ +0.24}$ & $=10.8$ \\

    &         &      & 0-260     & 260     & $24.1$ & $~11.4$ & $^{ +1.06 ~ +2.91 ~ +0.68 ~ +0.00}$ & $=15.1$ \\
    & & & & & & & & \\

h   & 150.3   &  17  & 0-350     & 350     & $30.2$ & $~12.4$ & $^{ +1.78 ~ +3.16 ~ +0.74 ~ +0.00}$ & $=16.7$ \\
    & & & & & & & & \\

control freq. & 300.6   &  30  & 0-350     & 350     & $70.3$ & $~26.4$ & $^{ +4.46 ~ +6.75 ~ +1.58 ~ +0.00}$ & $=36.0$ \\
    & & & & & & & & \\

i   & 626.5   &  10  & 50-200    & 150     & $53.4$ & $~25.6$ & $^{ +1.76 ~ +6.56 ~ +1.54 ~ +0.00}$ & $=33.9$ \\

l   &         &      & 190-260   & 70      & $47.4$ & $~19.4$ & $^{ +1.23 ~ +4.97 ~ +1.17 ~ +0.00}$ & $=25.7$ \\

    &         &      & 0-260     & 260     & $60.1$ & $~28.2$ & $^{ +2.70 ~ +7.22 ~ +1.69 ~ +0.00}$ & $=37.6$ \\
    & & & & & & & & \\

control freq. & 1253.0  &  10  & 50-200    & 150     & $114$ & $~49.4$ & $^{ +4.10 ~ +12.64 ~ +2.96 ~ +0.00}$ & $=65.6$ \\

    &         &      & 190-260   & 70      & $89.0$ & $~30.6$ & $^{ +2.69 ~ +7.84 ~ +1.84 ~ +0.00}$ & $=40.7$ \\

    &         &      & 0-260     & 260     & $107$ & $~53.5$ & $^{ +4.50 ~ +13.71 ~ +3.21 ~ +0.00}$ & $=71.2$ \\
    & & & & & & & & \\

m   & 1837.0  &  10  & 230-245   & 15      & $94.7$ & $~34.6$ & $^{ +1.26 ~ +8.86 ~ +2.08 ~ +0.00}$  & $=45.6$ \\

    &         &      & 0-245     & 245     & $192$ & $~54.9$ & $^{ +11.72 ~ +14.05 ~ +3.29 ~ +0.00}$ & $=76.5$\\

\end{tabular}
\end{center}
\caption{List of frequencies and observation times used in this analysis with the corresponding results. The first column describes the addressed QPO observation, labeled by letters as they appear in Tab. \ref{tab:tableofqpos}. A wider range of the detector's sensitivity can be explored using the frequency bands here labeled as control frequencies (see text). The second, third, fourth and fifth columns indicate the center frequency, bandwidth, interval, and duration used in the search. The sixth column provides the non-detection threshold. The last column presents the results where the contributions due to the different uncertainties are shown separately. The first two numbers in superscript represent the statistical uncertainty in the off-source estimation and calibration procedure respectively. The third one shows the contribution of a systematic uncertainty of $6\%$ due to the calibration procedure. The last uncertainty is a systematic arising from the off-source data modeling which depends on the presence of outliers (see text for details). To produce the upper bound $\mathrm{h}_\mathrm{rss-det}^{90\%}$ statistical contributions are added in quadrature while the systematic contributions are added linearly.}
\label{tab:upperbounds}
\end{table*}

Table \ref{tab:upperbounds} presents the results of this search, for both the control and QPO frequencies, in terms of $90\%$ upper bounds on the GW waveform strength, $\mathrm{h}_\mathrm{rss-det}^{90\%}$, measured at the time of the observation. The first column of the table indicates the observation we address, with reference to the original measurements shown in Tab. \ref{tab:tableofqpos}. The second, third, fourth and fifth columns indicate the center frequency, bandwidth, period, and duration used in the search. The sixth column, labeled as non-detection threshold, lists the maximum upper bound allowed in the non-detection regime. A data quality flag was used for the $92.5 ~ \mathrm{Hz}$ QPO observation only, with a power threshold set at the $2 \sigma$ level relative to tiles $125 ~ \mathrm{ms}$ long.

The last column, labeled $\mathrm{h}_\mathrm{rss-det}^{90\%}$, presents the results where the contributions due to the different uncertainties are shown separately. The first of these, the first number in superscript, shows the $90\%$ upper bound arising from the statistical uncertainties in the off-source estimation. These uncertainties are generated using a Monte-Carlo simulation: a set of means $\hat{\mu}$ and standard deviations $\hat{\sigma}$ are extracted from Gaussian distributed populations of standard deviation $\sigma_{\hat{\mu}}$ and $\sigma_{\hat{\sigma}}$ corresponding to the fit parameter uncertainties. For each $(\hat{\mu}, \hat{\sigma})$ combination and the same on-source excess power measure we used the lookup table in Ref. \cite{feldman98} to generate $90\%$ confidence intervals for the quoted upper limit.

The second uncertainty quoted is statistical and arises from errors in the detector response function to GW radiation via the calibration procedure. We placed a conservative estimate of the calibration accuracy to a one standard deviation of $20\%$. The third uncertainty is a systematic error of $6\%$ also arising from the calibration procedure.

The occasional presence of tails in the off-source segments, consisting typically of a few large excess power measurements in the off-source data of each QPO introduces a bias in the upper bounds which is presented as a source of systematic uncertainty (represented by the fourth number in superscript). This bias is quantified by including and excluding the off-source distribution $\pm 3 \sigma$ outliers from the fitting procedure and the difference in the upper bounds, $\delta h_\mathrm{rss-det}^{\mathrm{syst}} = h_\mathrm{rss-det}^\mathrm{with} - h_\mathrm{rss-det}^\mathrm{without}$ is shown in the column in question.

In order to fold in the different uncertainties we sum in quadrature the statistical uncertainties shown (originating from the off-source estimation and the calibration) and we increase the bound by the two systematic errors.

\section{Astrophysical Interpretation}
\label{sec:astrointerpreation}

In this section we provide a characteristic GW energy $E^{\mathrm{iso}}_\mathrm{GW}$ associated with the measured upper bounds $h_\mathrm{{rss-det}}^{90\%}$, shown in Tab. \ref{tab:upperbounds}, cast in terms of a simple source model. In this model we assume that the emission is isotropic, that the plus and cross polarization states are uncorrelated but have equal power.

Under these assumptions (equal uncorrelated power radiated in the plus and cross polarizations) the strain in the detector can be related to the GW flux incident on the Earth via
\begin{equation}
h_\mathrm{rss-det}^2 = \frac{1}{2} (F_+^2 + F_\times^2) h_\mathrm{rss}^2
\end{equation}
where
\begin{equation}
h_\mathrm{rss}^2 = \int_{-\infty}^\infty [ h_+^2(t) + h_\times^2(t) ] dt
\end{equation}
and $F_+$ and $F_\times$ are antenna response functions that depend on (i) the right-ascension and declination of the source, (ii) the time of the flare, (iii) the location and orientation of the detector, and (iv) a polarization angle defining the plus and cross polarizations. The dependence on this polarization angle vanishes in the combination $F_+^2 + F_\times^2$, which is a quantity ranging from 0 to 1; the Hanford detector's antenna response to SGR $1806-20$ at the time of the hyperflare was
\begin{equation}
F_+^2 + F_\times^2 = 0.174
\end{equation}
This shows that the source was not particularly well situated in the detectors antenna pattern. Under our assumption of isotropic emission, the energy released by the source is related to the gravitational wave flux at the Earth by
\begin{equation}
E^\mathrm{iso}_\mathrm{GW} = \frac{\pi^2 c^3 r^2 f_\mathrm{qpo}^2}{G} ~ h_\mathrm{rss}^2
\end{equation}
In terms of the upper limits presented, the equivalent bound on the gravitational wave emission corresponding to a particular QPO is
\begin{eqnarray}
E^{\mathrm{iso},90\%}_\mathrm{GW} = 4.29 \times 10^{-8} \mathrm{M}_\odot c^2 & \times & \\ \nonumber & & \hskip -5cm \left( \frac{r}{10 \mathrm{kpc}} \right)^2 \left( \frac{f_\mathrm{qpo}}{92.5 \mathrm{Hz}} \right)^2 \left( \frac{h_\mathrm{rss-det}^{90\%}} {4.53 \times 10^{-22} \mathrm{strain ~ Hz}^{-1/2}} \right)^2
\end{eqnarray}
(here the values of the best QPO strain bound are used). It is worth noting that the best energy upper bound is comparable to the energy emitted in the electromagnetic spectrum (see for example Ref. \cite{hurley05}).

\section{Conclusion}

Quasi-Periodic Oscillations have been observed in the pulsating X-ray tail of the SGR $1806-20$ hyperflare of 27 December 2004 by the RXTE and RHESSI satellites. The present consensus interprets the event as a dramatic re-configuration of the star's crust and/or magnetic field. In turn, this {\it starquake} could plausibly excite the star's global seismic modes and the observed QPOs could potentially be driven by the seismic modes. The energetics of the event, the close proximity of the source, and the availability of observed QPO frequencies and bandwidths provided a unique opportunity to measure GWs associated with this phenomenon.

Upper limits in the gamma and high-energy neutrino flux were recently measured by the AMANDA-II detector \cite{amanda06}. However the only other published GW search associated with the SGR $1806-20$ hyperflare used the AURIGA bar detector \cite{auriga05} to place upper limits on the GW waveform strength emitted for frequencies around $\sim 900 ~ \mathrm{Hz}$. At the time of the event, H1's strain noise equivalent in the $\sim 900 ~ \mathrm{Hz}$ region is a factor $\sim 5$ lower than AURIGA's.

The AURIGA search targeted different physics, therefore the comparison to our results is not possible. Exponentially decaying sinusoids of decay time $100 ~ \mathrm{ms}$ were searched for by measuring the power in time and frequency slices of $\Delta t = 201.5 ~ \mathrm{ms}$ and $\Delta f = 5 ~ \mathrm{Hz}$ respectively in the $855 ~ \mathrm{Hz}$ to $945 ~ \mathrm{Hz}$ range. A set of $95\%$ upper bounds on the waveform strength were placed in the $h_\mathrm{rss-det}^{95\%} = 1.4 \times 10^{-21} ~ \mathrm{strain ~ Hz}^{-1/2}$ to $h_\mathrm{rss-det}^{95\%} = 3.5 \times 10^{-21} ~ \mathrm{strain ~ Hz}^{-1/2}$ range.

At the time of the event one of the three LIGO detectors was in operation under the {\it Astrowatch} program. Under this program, data is collected at times of commissioning when the interferometers are not undergoing adjustments. Only $\sim 2 ~ \mathrm{h}$ of data was available for this analysis.

An algorithm was designed to measure the {\it excess power} deposited in the machine at the time of the event. This algorithm exploits power measures in multiple bands to reject common mode noise sources, such as broadband noise. Power measures in time scales less than 1s are also monitored to reject {\it fast} signatures inconsistent with the scope of this analysis.

The design was driven by the desire to repeat this measurement for future flares with the ability to use multiple data streams from multiple detectors, focusing on modularity, flexibility, and simplicity.

Signals were software-injected into the raw data stream to study the analysis sensitivity to a variety of waveform families and parameters. A large astrophysical motivated parameter space was explored under which the search sensitivity is essentially constant.

At the time of the event, the strain-equivalent amplitude spectral density of the detector output was a factor of a few away from the one corresponding to the fourth science run. Under this condition, the best upper limit that we place corresponds to the $92.5 ~ \mathrm{Hz}$ QPO observed $150 ~ \mathrm{s}$ to $260 ~ \mathrm{s}$ seconds after the flare. In terms of waveform strength, we place a $90\%$ upper bound of $h_\mathrm{rss-det}^{90\%} = 4.53 \times 10^{-22} ~ \mathrm{strain ~ Hz}^{-1/2}$ on the GW waveform strength in the detectable polarization state reaching our Hanford (WA) detector, which, in terms of a simple source model, provides a characteristic energy $E_{\mathrm{GW}}^{\mathrm{iso,90\%}} = 7.67 \times 10^{46} ~ \mathrm{erg} ~ (4.29 \times 10^{-8} ~ \mathrm{M}_{\odot} \mathrm{c}^2)$. This is the best upper limit published on the GW waveform strength on this type of source and represents the first multiple-frequency asteroseismology measurement using a GW detector. It is also worth noting that this energy estimate is of the same order as the isotropic energy estimate measured electromagnetically, providing the opportunity to probe the energy reservoir of the source.

The limits presented here represent GW strength obtained by the LIGO detectors in late 2004. At the time of this writing, LIGO is undergoing a data-taking period, referred to as the fifth science run S5, where all three interferometers have reached design sensitivity, \cite{S5}. The improvement at $150 ~ \mathrm{Hz}$ corresponds to a decrease in strain-equivalent noise of $\geq 3$ in terms of GW energetics. This estimate excludes the sensitivity increase that can be achieved by cross-correlating data streams from the multiple LIGO detectors. A follow-up of this analysis will certainly examine the various SGR $1806-20$/SGR $1900+14$ outbursts, which occurred in the 2005 - 2006 period, exploring GW energetics which probe the $\sim 2 \times 10^{45} ~ \mathrm{erg} ~ ( \sim 10^{-9} ~ \mathrm{M}_{\odot} \mathrm{c}^2)$ regime.

At the end of the S5 data-taking period, the initial LIGO detectors will be upgraded to an enhanced state \cite{enhancedLIGO} which we refer to as Enhanced LIGO. The foreseen improvement will be a factor of $\sim 2$ in strain-equivalent noise for frequencies above $100 ~ \mathrm{Hz}$. The future GEO-HF \cite{geohf} detector will provide a significant high-frequency improvement in sensitivity providing an opportunity to study future high-frequency QPOs.

Advanced LIGO \footnote{http://www.ligo.caltech.edu/advLIGO/} will provide an increase in strain-equivalent sensitivity of $\sim 10$ with respect to the initial LIGO detectors while opening up the low ($10 - 50 ~ \mathrm{Hz}$) frequency range. This offers a particularly interesting opportunity because a lower frequency search would be feasible. For hyperflare events occurring at the time of its operation, the observable GW energetics at $100 ~ \mathrm{Hz}$ would lie in the  $\sim 2 \times 10^{43} ~ \mathrm{erg} ~ ( \sim 10^{-11} ~ \mathrm{M}_{\odot} \mathrm{c}^2)$ regime.

\section{Acknowledgments}

We are indebted to Gianluca Israel and Anna Watts for frequent and fruitful discussions. The authors gratefully acknowledge the support of the United States National Science Foundation for the construction and operation of the LIGO Laboratory and the Particle Physics and Astronomy Research Council of the United Kingdom, the Max-Planck-Society and the State of Niedersachsen/Germany for support of the construction and operation of the GEO600 detector. The authors also gratefully acknowledge the support of the research by these agencies and by the Australian Research Council, the Natural Sciences and Engineering Research Council of Canada, the Council of Scientific and Industrial Research of India, the Department of Science and Technology of India, the Spanish Ministerio de Educacion y Ciencia, The National Aeronautics and Space Administration, the John Simon Guggenheim Foundation, the Alexander von Humboldt Foundation, the Leverhulme Trust, the David and Lucile Packard Foundation, the Research Corporation, the Alfred P. Sloan Foundation and Columbia University in the City of New York.

\bibliography{References}

\end{document}

%% file: LSCauthors.tex
%
%
%
\affiliation{Albert-Einstein-Institut, Max-Planck-Institut f\"ur Gravitationsphysik, D-14476 Golm, Germany}
\affiliation{Albert-Einstein-Institut, Max-Planck-Institut f\"ur Gravitationsphysik, D-30167 Hannover, Germany}
\affiliation{Andrews University, Berrien Springs, MI 49104 USA}
\affiliation{Australian National University, Canberra, 0200, Australia}
\affiliation{California Institute of Technology, Pasadena, CA  91125, USA}
\affiliation{Caltech-CaRT, Pasadena, CA  91125, USA}
\affiliation{Cardiff University, Cardiff, CF2 3YB, United Kingdom}
\affiliation{Carleton College, Northfield, MN  55057, USA}
\affiliation{Charles Sturt University, Wagga Wagga, NSW 2678, Australia}
\affiliation{Columbia University, New York, NY  10027, USA}
\affiliation{Embry-Riddle Aeronautical University, Prescott, AZ   86301 USA}
\affiliation{Hobart and William Smith Colleges, Geneva, NY  14456, USA}
\affiliation{Inter-University Centre for Astronomy  and Astrophysics, Pune - 411007, India}
\affiliation{LIGO - California Institute of Technology, Pasadena, CA  91125, USA}
\affiliation{LIGO Hanford Observatory, Richland, WA  99352, USA}
\affiliation{LIGO Livingston Observatory, Livingston, LA  70754, USA}
\affiliation{LIGO - Massachusetts Institute of Technology, Cambridge, MA 02139, USA}
\affiliation{Louisiana State University, Baton Rouge, LA  70803, USA}
\affiliation{Louisiana Tech University, Ruston, LA  71272, USA}
\affiliation{Loyola University, New Orleans, LA 70118, USA}
\affiliation{Moscow State University, Moscow, 119992, Russia}
\affiliation{NASA/Goddard Space Flight Center, Greenbelt, MD  20771, USA}
\affiliation{National Astronomical Observatory of Japan, Tokyo  181-8588, Japan}
\affiliation{Northwestern University, Evanston, IL  60208, USA}
\affiliation{Rochester Institute of Technology, Rochester, NY 14623, USA}
\affiliation{Rutherford Appleton Laboratory, Chilton, Didcot, Oxon OX11 0QX United Kingdom}
\affiliation{San Jose State University, San Jose, CA 95192, USA}
\affiliation{Southeastern Louisiana University, Hammond, LA  70402, USA}
\affiliation{Southern University and A\&M College, Baton Rouge, LA  70813, USA}
\affiliation{Stanford University, Stanford, CA  94305, USA}
\affiliation{Syracuse University, Syracuse, NY  13244, USA}
\affiliation{The Pennsylvania State University, University Park, PA  16802, USA}
\affiliation{The University of Texas at Brownsville and Texas Southmost College, Brownsville, TX  78520, USA}
\affiliation{Trinity University, San Antonio, TX  78212, USA}
\affiliation{Universitat de les Illes Balears, E-07122 Palma de Mallorca, Spain}
\affiliation{Universit\"at Hannover, D-30167 Hannover, Germany}
\affiliation{University of Adelaide, Adelaide, SA 5005, Australia}
\affiliation{University of Birmingham, Birmingham, B15 2TT, United Kingdom}
\affiliation{University of Florida, Gainesville, FL  32611, USA}
\affiliation{University of Glasgow, Glasgow, G12 8QQ, United Kingdom}
\affiliation{University of Maryland, College Park, MD 20742 USA}
\affiliation{University of Michigan, Ann Arbor, MI  48109, USA}
\affiliation{University of Oregon, Eugene, OR  97403, USA}
\affiliation{University of Rochester, Rochester, NY  14627, USA}
\affiliation{University of Salerno, 84084 Fisciano (Salerno), Italy}
\affiliation{University of Sannio at Benevento, I-82100 Benevento, Italy}
\affiliation{University of Southampton, Southampton, SO17 1BJ, United Kingdom}
\affiliation{University of Strathclyde, Glasgow, G1 1XQ, United Kingdom}
\affiliation{University of Washington, Seattle, WA, 98195}
\affiliation{University of Western Australia, Crawley, WA 6009, Australia}
\affiliation{University of Wisconsin-Milwaukee, Milwaukee, WI  53201, USA}
\affiliation{Washington State University, Pullman, WA 99164, USA}
\author{B.~Abbott}\affiliation{LIGO - California Institute of Technology, Pasadena, CA  91125, USA}
\author{R.~Abbott}\affiliation{LIGO - California Institute of Technology, Pasadena, CA  91125, USA}
\author{R.~Adhikari}\affiliation{LIGO - California Institute of Technology, Pasadena, CA  91125, USA}
\author{J.~Agresti}\affiliation{LIGO - California Institute of Technology, Pasadena, CA  91125, USA}
\author{P.~Ajith}\affiliation{Albert-Einstein-Institut, Max-Planck-Institut f\"ur Gravitationsphysik, D-30167 Hannover, Germany}
\author{B.~Allen}\affiliation{Albert-Einstein-Institut, Max-Planck-Institut f\"ur Gravitationsphysik, D-30167 Hannover, Germany}\affiliation{University of Wisconsin-Milwaukee, Milwaukee, WI  53201, USA}
\author{R.~Amin}\affiliation{Louisiana State University, Baton Rouge, LA  70803, USA}
\author{S.~B.~Anderson}\affiliation{LIGO - California Institute of Technology, Pasadena, CA  91125, USA}
\author{W.~G.~Anderson}\affiliation{University of Wisconsin-Milwaukee, Milwaukee, WI  53201, USA}
\author{M.~Arain}\affiliation{University of Florida, Gainesville, FL  32611, USA}
% ...

\collaboration{The LIGO Scientific Collaboration, http://www.ligo.org}
\noaffiliation